\documentclass[aps,prx,amsmath,amssymb,reprint,nofootinbib]{revtex4-1}\usepackage{graphicx}
\usepackage{overpic}
\usepackage{dashbox}
\usepackage{afterpage}
\usepackage[dvipsnames]{xcolor}
\usepackage[caption=false]{subfig}
\usepackage{MnSymbol}
\usepackage{mathtools}
\newcommand\phiinket{|\Phi_{\textrm{in}}\rangle}
\newcommand\phiinbra{\langle\Phi_{\textrm{in}}|}
\newcommand\phioutket{|\Phi_{\textrm{out}}\rangle}

\newcommand\phioutbra{\langle\Phi_{\textrm{out}}|}
\newcommand\DG{\langle\Delta\Gamma\rangle}
\newcommand\DGS{\langle\Delta\Gamma(\Smat)\rangle}
\newcommand\DGSpar{\langle\Delta\Gamma(\Smatpar)\rangle}

\newcommand\etagammain{|\eta \ \gamma\rangle^{\textrm{in}}}

\newcommand\etagammaout{|\eta \ \gamma\rangle^{\textrm{out}}}
\newcommand\etagammainout{|\eta \ \gamma\rangle^{\textrm{in(out)}}}
\newcommand\baretagammainout{{}^{\textrm{in(out)}}\langle\bar{\gamma} \ \bar{\eta}|}
\newcommand\baretagammaout{{}^{\textrm{out}}\langle\bar{\gamma} \ \bar{\eta}|}

\newcommand\Dabs{\frac{A\Gamma+\Gamma A}{2}}
\newcommand\Dasy{\frac{S^{\dagger}[S,\Gamma]+[S,\Gamma]^{\dagger}S}{2}}
\newcommand\Dabsmat{\mat{A}\Gammamat+\Gammamat \mat{A}}
\newcommand\Dasymat{\Smat^{\dagger}[\Smat,\Gammamat]+[\Smat,\Gammamat]^{\dagger}\Smat}

\newcommand\ain{\alpha_{\gamma}^{\eta}}
\newcommand\aout{\beta_{\gamma}^{\eta}}
\newcommand\ainvec{{\bunderline{\alpha}}}
\newcommand\aoutvec{{\bunderline{\beta}}}

\newcommand{\bunderline}[1]{\underline{#1\mkern-4mu}\mkern4mu }
\newcommand\scatt{\rho_{\gamma}^{\eta}}
\newcommand\scattvec{\bunderline{\rho}}
\newcommand\Tmat{\bunderline{\bunderline{T}}}

\newcommand\inc{\mu_{\gamma}^{\eta}}
\newcommand\incvec{\bunderline{\mu}}
\newcommand\Gammamat{\bunderline{\bunderline{\Gamma}}}
\newcommand\Smat{\bunderline{\bunderline{S}}}
\newcommand\Smatpar{\bunderline{\bunderline{S}}_{\Pi}}

\newcommand\Rmat{\bunderline{\bunderline{R}}}
\newcommand\Rmatinv{\bunderline{\bunderline{R}}^{-1}}
\newcommand\Rmata{\bunderline{\bunderline{R}}_1}
\newcommand\Rmatb{\bunderline{\bunderline{R}}_2}
\newcommand\Rmatbinv{\bunderline{\bunderline{R}}^{-1}_2}
\newcommand\Rmatainv{\bunderline{\bunderline{R}}^{-1}_1}
\newcommand\Mmata{\bunderline{\bunderline{M}}_1}
\newcommand\Mmatb{\bunderline{\bunderline{M}}_2}

\newcommand\mat[1]{\bunderline{\bunderline{#1}}}

\newcommand\vect[1]{\bunderline{#1}}

\newcommand\impliesdueto[1]{\stackrel{#1}{\implies}}
\newcommand\duetoeq[1]{\stackrel{Eq.\ (\ref{#1})}{=}}
\newcommand\dboxed[1]{\dbox{\ensuremath{#1}}}

\begin{document}
\title{Unified theory to describe and engineer conservation laws in light-matter interactions}
\author{Ivan Fernandez-Corbaton}
\email{ivan.fernandez-corbaton@kit.edu}
\affiliation{Institute of Nanotechnology, Karlsruhe Institute of Technology, 76021 Karlsruhe, Germany}
\author{Carsten Rockstuhl}\affiliation{Institut f\"ur Theoretische Festk\"orperphysik, Karlsruhe Institute of Technology, 76131 Karlsruhe, Germany}
\affiliation{Institute of Nanotechnology, Karlsruhe Institute of Technology, 76021 Karlsruhe, Germany}
\begin{abstract}
	The effects of the electromagnetic field on material systems are governed by joint light-matter conservation laws. An increasing number of these balance equations are currently being considered both theoretically and with an eye to their practical applicability. We present a unified theory to treat conservation laws in light-matter interactions. It can be used to describe and engineer the transfer of any measurable property from the electromagnetic field to any object. The theory allows to explicitly characterize and separately compute the transfer due to asymmetry of the object and the transfer due to field absorption by the object. It also allows to compute the upper bound of the transfer rate of any given property to any given object, together with the corresponding most efficient illumination which achieves the bound. Due to its algebraic nature, the approach is inherently suited for computer implementation.
\end{abstract}
\maketitle
\section{Introduction and summary}
Properties like energy, linear momentum, and angular momentum can be transferred from the electromagnetic field to material systems during light-matter interactions. These exchanges are governed by conservation laws that apply to the combined system of field and matter. These joint conservation laws are, arguably, among the most important theoretical principles in electrodynamics. Some of them have been put to practical use for some time now. For example, the exchange of linear momentum and angular momentum give rise to optical forces and torques. This allows the optical manipulation of objects, like in optical trapping and optical tweezers applications. The practical use of other less well known conservation laws is being investigated. For example, the helicity conservation law is being considered in the context of chiral light-matter interactions \cite{Nieto2015,Nienhuis2016,Poulikakos2016,Zambrana2016b}. The effects of the conservation laws of the two transverse components of angular momentum are also under scrutiny \cite{Andrews2004,Albaladejo2009,Iglesias2011,Nieto2015b,Nienhuis2016}. One of the areas of application of these less well known conservation laws is envisioned to be the manipulation of chiral matter with chiral light \cite{Hakobyan2014,Tkachenko2014,Cameron2014,Tkachenko2014b,Canaguier2013,Hayat2015,Rahimzadegan2016b}.

In this article, and motivated by the growing number of interesting conservation laws, we develop a unified theory for the analysis and engineering of the transfer of properties from the electromagnetic field to material systems. Any measurable property can be treated in the same way. 

In Sec. \ref{sec:setting} we introduce the setting and mathematical tools that we use in the article. We also outline and discuss the assumptions that we make. In Sec. \ref{sec:absvsasym} we identify a structure underlying all conservation laws: The total transfer of any measurable property can be divided into a part due to absorption of the field by the object and a part due to asymmetry of the object. One is zero if the object is non-absorbing. The other is zero if the object has the symmetry related to the property in question. We provide the expressions of the operators that determine both kinds of transfer, and discuss them in relation to Noether's theorem \cite{Noether1918}. After introducing bases for the incoming and outgoing fields in Sec. \ref{sec:basis}, we obtain a universal equation for the transfer of any property. The form of the equation does not depend on the property. We then derive expressions for the transfer rate per Watt of incoming power in Sec. \ref{sec:transferrates}, and show their suitability for engineering purposes in Sec. \ref{sec:engineering}. For example, given an object, the expressions that we obtain allow to compute both the maximum achievable transfer rate per incoming Watt, and the exact incoming field which achieves such upper bound. We also discuss other engineering possibilities. In Sec. \ref{sec:dtransfer} we provide the means to compute optimal incoming fields for chiral sorting. In Sec. \ref{sec:implementation} we develop the formalism further into a computer friendly formulation and  provide guidance for its implementation using T-matrix algorithms. In Sec. \ref{sec:examples}, we show two examples of application. In Sec. \ref{sec:example1}, the transfer rates of different quantities between a plane-wave and a complicated composite object are rigorously obtained. We analyze the transfer of energy, linear and angular momentum, helicity, and transverse ``spin angular momentum''. We quantitatively show the rate of transfer due to absorption and that due to asymmetry in each case, and relate the latter to the symmetries of the object. We then show a comparison between the force and helicity transfer under plane-wave illumination, and the maximum that can be achieved in each case with a monochromatic beam. Section \ref{sec:example2} contains an example of application of the chiral sorting theory developed in Sec. \ref{sec:dtransfer}. Finally, Sec. \ref{sec:conclusions} contains the conclusions.

\section{Setting\label{sec:setting}}
\begin{figure}[h!]
	\includegraphics[width=\linewidth]{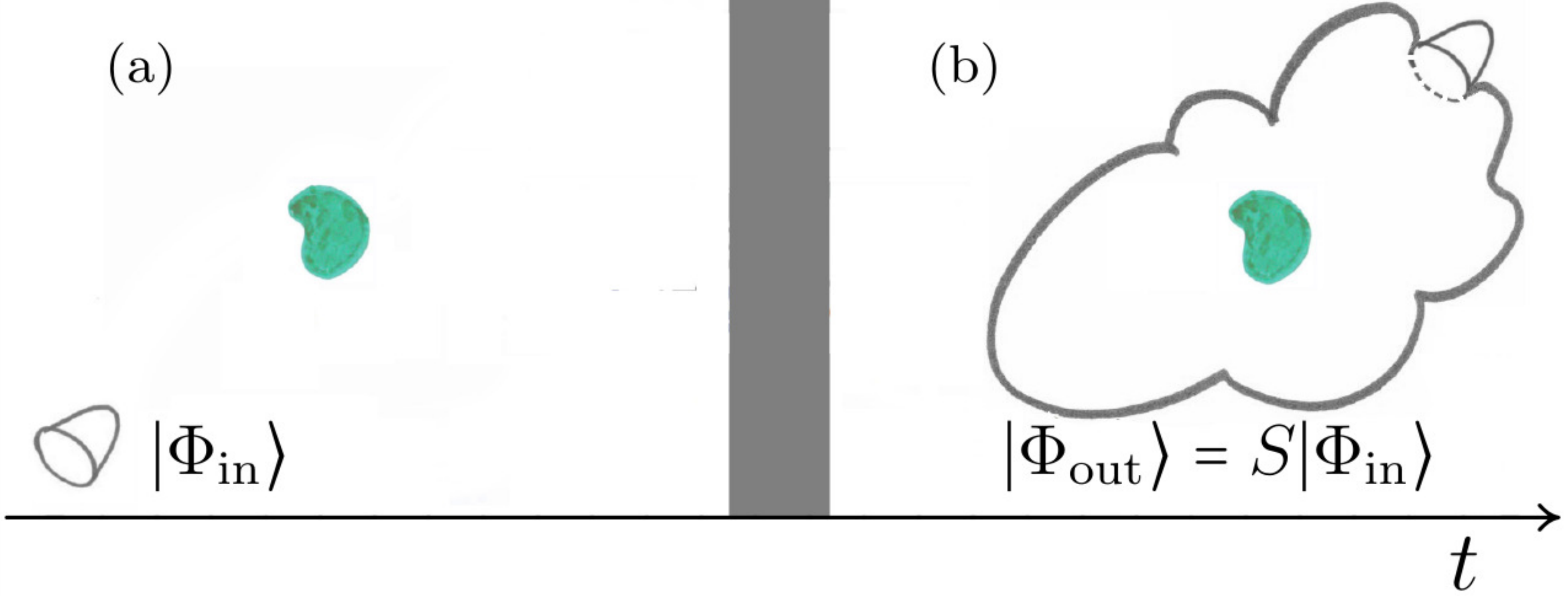}
										\caption{ \label{fig:scatt_op} Successive phases of the light-matter interaction. The bullet-shaped field in the incoming phase (a), and the cloud-plus-bullet-shaped field in the outgoing phase (b) do not interact with the object. The interaction time is confined to the grayed-out region. The object is represented by the bean-shaped drawings.  The scattering operator of the object ($S$) relates the incoming and outgoing fields $\phioutket=S\phiinket$.}
\end{figure}

The light-matter interaction setting is sketched in Fig. \ref{fig:scatt_op}. In a first phase depicted in Fig. \ref{fig:scatt_op}(a), the field and the object do not interact. In a second phase, gray in the figure, they interact during some finite amount of time. In a third phase depicted in Fig. \ref{fig:scatt_op}(b), the field and the object are again impervious to each other. From the point of view of the object, the electromagnetic field is incoming during the first phase and outgoing during the third phase. 

In this setting, the exact description of what happens during the grayed area is not used. All the information about the interaction processes that has an observable physical meaning is contained in the mapping from the incoming to the outgoing fields. This is the central idea in Heisenberg's scattering matrix formulation of quantum mechanics \cite[p. 261, p. 321]{Van2007}. It is still relevant in current physical theories (\cite[\S 19 p. 314]{Birula1975}\cite[\S 1, p. 3]{Berestetskii1982}, \cite[Chap. 12]{Weinberg1995}, \cite[Preamble, Chap. 2 p. 45]{Scharf2014}).

Let us now consider a measurable property like energy, momentum, angular momentum, helicity, etc ... subject to a conservation law. We call this generic property $\Gamma$. We can quantify the transfer of $\Gamma$ between the field and the object by measuring $\Gamma$ in the incoming field, measuring $\Gamma$ in the outgoing field, and subtracting the two. To do so, we must specify how to measure $\Gamma$ in the incoming and outgoing fields. To this end, we introduce the mathematical tools that we are going to use in this article, namely Hilbert spaces. We will use Dirac's ``$\langle$bra$|$'' ``$|$ket$\rangle$'' notation\footnote{Dirac introduced this notation in quantum mechanics. It is also very convenient for other situations that can be treated using the framework of Hilbert spaces.}.

In this description, the fields are vectors in the incoming and outgoing Hilbert spaces of transverse solutions to Maxwell's equations: $\phiinket$ and $\phioutket$. Classical real fields are conveniently mapped onto the complex Hilbert spaces through their complex representation, which uses only the positive frequencies of the field's harmonic decomposition (see e.g. \cite[Chap. 3.1]{Mandel1995} or \cite[Chap. 10.2]{Born1999}).

The effect of the object on the fields is represented by the linear scattering operator $S$. It maps incoming fields onto outgoing fields: 
\begin{equation}
	\label{eq:outSin}
	\phioutket=S\phiinket.
\end{equation}
In this article, we restrict ourselves to linear scatterers. The property $\Gamma$ is also represented by a linear operator.

We will measure the transfer of $\Gamma$ from light to matter during the interaction by subtracting the integrated value of $\Gamma$ in the outgoing state from that in the incoming state\footnote{The integrated value of an operator $X$ for the vector $|\Psi\rangle$ is obtained by the scalar product of the vectors $X|\Psi\rangle$ and $|\Psi\rangle$: $\langle\Psi|X|\Psi\rangle$. As explained in \cite[Sec. 5.2]{Birula1996}, \cite[Sec. 3.2]{FerCorThesis} and \cite[Sec. III]{FerCor2012b}, the integrated values of operators like energy, momentum, angular momentum, etc ..., computed as $\langle\Psi|X|\Psi\rangle$, coincide with the results of the typical spatial integrals involving the electromagnetic fields: For example $c/(4\pi)\int d\mathbf{r}\ \mathbf{D}\times \mathbf{B}$ for the linear momentum.}. The difference between the two integrated values 
\begin{equation}
	\label{eq:transfer}
	\DG=\phiinbra\Gamma\phiinket-\phioutbra\Gamma\phioutket
\end{equation}
must represent the amount of $\Gamma$ that has been transferred from light to matter during the interaction.

Using the Hermitian conjugate version of Eq. (\ref{eq:outSin}) [$\phioutbra=\phiinbra S^\dagger$], Eq. (\ref{eq:transfer}) can be written as
\begin{equation}
	\label{eq:transfer_dev}
	\begin{split}
		\DG&=\phiinbra\Gamma\phiinket-\phiinbra S^{\dagger}\Gamma S\phiinket\\
		   &=\phiinbra\Gamma-S^{\dagger}\Gamma S\phiinket.
	\end{split}
\end{equation}
We now impose that $\DG$ must be always real for any $\phiinket$. This forces $\Gamma-S^{\dagger}\Gamma S$ to be a Hermitian operator, which then forces $\Gamma$ to be a Hermitian operator as well: $\Gamma=\Gamma^{\dagger}$. This is the case for all the conservation laws that we are aware of. For example, it is the case for the most commonly considered conservation laws: Energy, momentum, and angular momentum. It is also the case for the conservation laws of helicity and the two parts of the transverse split of angular momentum $\mathbf{J}=\mathbf{\hat{L}}+\mathbf{\hat{S}}$ \cite{Drummond1999,Drummond2006,Cameron2012,Bliokh2013,Nieto2015,Nieto2015b}. 

We will call the properties for which $\DG$ is real valued for all $\phiinket$ {\em measurable} properties. The motivation for this naming is the following: Any measurement device can be thought of as a material system interacting with the electromagnetic field. The measurement device is triggered by the exchange of properties like energy and momentum. It seems physically justified to restrict the measurable properties of the field to those in which the exchange $\DG$ is real valued. For example, in the transfer of linear and angular momentum, the restriction corresponds to the physical fact that forces and torques are real valued. 

In the interest of conciseness, we will from now on write {\em property} as to mean {\em measurable property represented by a Hermitian operator}.

The idea of subtracting incoming and outgoing fluxes is used for the linear and angular momentum in \cite{Nieminen2007,Nieminen2014}. This section generalizes its application to any property.

A note about the applicability of Eqs. (\ref{eq:transfer})-(\ref{eq:transfer_dev}) is now in order.
\subsection{Steady state\label{sec:steadystate}}
The incoming field depicted in Fig. \ref{fig:scatt_op} (a) can be referred to as a wave packet. If, instead of well separated wave packets, the illumination consists of a continuous beam like a CW laser, the above methodology can be used to compute the {\em transfer rate} of $\Gamma$ (see Sec. \ref{sec:transferrates}). The transfer rates are meaningful if the system is in a steady state with respect to the incoming and outgoing electromagnetic fluxes. After the continuous beam source is turned on, a steady state is quickly reached for the incoming flux. The same is true for the outgoing flux as long as the response of the object does not change significantly. This last assumption is only going to hold temporarily. For example, the interaction with the field may eventually cause the object to change position and orientation. Then, the expression of the operator $S$ changes accordingly: It must be translated and rotated. Changes that are more substantial than translations and rotations may also happen, for example, if the field deforms the object significantly. In many situations we expect that the time scale in which $S$ changes will be much longer than the time needed for reaching a steady state in the outgoing flux. As long as the two time scales are distinct enough, a piece-wise steady state can be assumed and the methodology applied with a constant $S$. Then, after some period of time, a new $S$ operator may be considered taking into account the changes that the field has produced in the object, and so on. 

The time scale for reaching a steady outgoing flux is set by what has been called interaction time \cite{Wigner1955,Smith1960}. It also determines the length of the gray region in Fig. \ref{fig:scatt_op}. It is very often on the order of a few tens of attoseconds \cite{Pazourek2015}. In extreme cases it may reach tens of nanoseconds when metastable resonant states in the matter are excited by the incoming fields \cite{Bourgain2013}.
\section{Transfer due to absorption versus transfer due to asymmetry\label{sec:absvsasym}}
We now return to Eq. (\ref{eq:transfer_dev}). We see that when
\begin{equation}
	\label{eq:cons}
	\Gamma=S^\dagger \Gamma S,
\end{equation}
the transfer of $\Gamma$ between light and matter is forbidden. That is: $\DG=0$ for all possible illuminations $\phiinket$. 

We will now investigate the relationship between the $\DG=0$ condition and the symmetries of the object. 

One salient property of Hermitian operators is that their exponentiation generates unitary transformations parametrized by a real number $\theta$ 
\begin{equation}
	\label{eq:genT}
	T_\Gamma(\theta)=\exp\left(-i\theta\Gamma\right)=\sum_{l=0}^{\infty} \frac{\left(-i\theta\Gamma\right)^l}{l!}.
\end{equation}
These transformations are often physically relevant. For example, energy generates time translations, momentum generates spatial translations, angular momentum generates rotations, and helicity generates electromagnetic duality. The transformations generated by the two parts of the transverse split of the angular momentum $\mathbf{J}=\mathbf{\hat{L}}+\mathbf{\hat{S}}$ can be found in \cite{FerCor2013b}. In particular, $\mathbf{\hat{S}}$ generates helicity and frequency dependent translations.

An object is invariant under a given transformation when the commutator between $T_\Gamma(\theta)$ and the scattering operator of the object vanishes: 
\begin{equation}
	\label{eq:commutT}
	[T_\Gamma(\theta),S]=T_\Gamma(\theta)S-ST_\Gamma(\theta)=0.
\end{equation}
We say that the object has the symmetry $T_\Gamma(\theta)$. It can be shown\footnote{In order to show that Eq. (\ref{eq:commut}) implies Eq. (\ref{eq:commutT}), one uses Eq. (\ref{eq:genT}) and the definition of commutator $[A,B]=AB-BA$. The proof that Eq. (\ref{eq:commutT}) implies Eq. (\ref{eq:commut}) can be found in \cite[Sec. 8.1.1]{Elbaz1998}.} that Eq. (\ref{eq:commutT}) is equivalent to the vanishing of the commutator between $S$ and the generator $\Gamma$. 
\begin{equation}
	\label{eq:commut}
	[\Gamma,S]=0,\text{ i.e. }\Gamma S=S\Gamma.
\end{equation}
For example, if $S$ commutes with the linear momentum in the $x$ direction ($[P_x,S]=0$), it implies that $[\exp\left(-i\theta P_x\right),S]=0$. This means that the response of the object has continuous translational symmetry in the $x$ direction, i.e. it is invariant under translations in the $x$ direction.

Equation (\ref{eq:commut}) is also equivalent\footnote{Showing that Eq. (\ref{eq:commut}) implies Eq. (\ref{eq:eigcons}) is straightforward. To show the converse implication, we use that $\Gamma$ can  be chosen to be one of the four commuting operators whose eigenvalues characterize a basis of incoming fields, as discussed in Sec. \ref{sec:basis}. We then expand the incoming field into such basis $\phiinket=\sumint_{\eta,\gamma}\ain\etagammain$ and apply the commutator operator $\Gamma S-S\Gamma$ to it. The first term is $\Gamma S\phiinket=\sumint_{\eta,\gamma}\ain\Gamma\left(S\etagammain\right)=\sumint_{\eta,\gamma}\ain\gamma\left(S\etagammain\right)$, where the last equality follows from assuming Eq. (\ref{eq:eigcons}); note that $S\etagammain$ is outgoing. The second term is $-S\Gamma\phiinket=-S\sumint_{\eta,\gamma}\gamma\ain\etagammain=-\sumint_{\eta,\gamma}\gamma\ain\left(S\etagammain\right)$. The sum of the two terms is zero, which means that Eq. (\ref{eq:commut}) is met.} to the following condition:
\begin{equation}
	\label{eq:eigcons}
\Gamma\phiinket=\gamma_0\phiinket\implies\Gamma\phioutket=\gamma_0\phioutket.
\end{equation}
Namely: If the incoming field is an eigenvector of $\Gamma$ with eigenvalue $\gamma_0$, the outgoing field will also be an eigenvector of $\Gamma$ with the same eigenvalue. This, however, does not necessarily imply that $\DG$ will be zero. The conditions in Eqs. (\ref{eq:cons}) and (\ref{eq:commut}) are not equivalent. It is straightforward to check that they become equivalent if there is neither absorption nor gain in the light-matter interaction \footnote{In this case, $S$ is unitary, which implies that it is invertible, and that $S^{-1}=S^\dagger$. Then, if Eq. (\ref{eq:cons}) is met, we can left multiply it by $S$ to get Eq. (\ref{eq:commut}). Similarly, left multiplying Eq. (\ref{eq:commut}) by $S^\dagger$ results in Eq. (\ref{eq:cons}).}.

We conclude that {\em objects that have the symmetry generated by $\Gamma$ can only exchange $\Gamma$ with the field if the interaction is not unitary}. These considerations suggest the possibility of writing $\DG$ as the sum of two contributions: One that is zero if the object has the symmetry generated by $\Gamma$, and another one that is zero if the interaction is unitary. We now perform such split, for which we first define an operator related to the (non-)unitary character of $S$.

When the interaction is not unitary, the norms of the incoming and outgoing fields need not be equal. We can define an operator $A$ which determines such norm changes:
\begin{equation}
	\begin{split}
		\phiinbra\Phi_{\textrm{in}}\rangle-\phioutbra\Phi_{\textrm{out}}\rangle\duetoeq{eq:outSin}&\phiinbra I-S^\dagger S\phiinket\\
		=&\phiinbra A \phiinket,
	\end{split}
\end{equation}
where the last equality is the definition:
\begin{equation}
	\label{eq:A}
	A=I-S^\dagger S.
\end{equation}
From now on, we will assume that there is no gain in the interaction, so that $A$ represents the absorption of the field by the object during the interaction. 

Together, the operators $S$ and $A$ allow for a straightforward interpretation of their role in light-matter interactions. This is achieved by the relationship between the singular value decompositions of the $S$ and $A$ operators established in App. \ref{app:svd}. A salient fact is that the right(left) singular vectors of $S$ form a basis for the incoming(outgoing) fields. Then, the energy of a general incoming field contained in each right singular vector is split by the object into two parts: One is absorbed by the object, the other goes to the outgoing field, but only into a single outgoing mode: The corresponding left singular vector. The ratio between re-radiated and absorbed energy is $(d_l)^2/(1-d_l^2)$, where $d_l$ is the corresponding singular value, which is always a real non-negative number. Note that no gain implies $d_l\le 1$, and neither gain nor loss implies $d_l=1$. Along these lines, we can also interpret $(1-d_l^2)/(d_l)^2$ as the proportion of energy that changes its character from propagating to localized.

To advance towards the split, we perform two manipulations of Eq. (\ref{eq:transfer_dev}). In the first one, we substitute $\Gamma S=S\Gamma-\left[S,\Gamma\right]$:  
\begin{equation}
	\label{eq:dgabssym1}
	\begin{split}
		\DG=&\phiinbra\Gamma-S^{\dagger}\dboxed{\Gamma S}\phiinket\\
			&=\phiinbra\Gamma-S^{\dagger}\dboxed{\left(S\Gamma-\left[S,\Gamma\right]\right)}\phiinket\\
		&=\phiinbra\Gamma-S^{\dagger}S\Gamma+S^{\dagger}\left[S,\Gamma\right]\phiinket\\
		&=\phiinbra\left(I-S^{\dagger} S\right)\Gamma\phiinket+\phiinbra S^{\dagger}\left[S,\Gamma\right]\phiinket\\
		&=\phiinbra A\Gamma\phiinket+\phiinbra S^{\dagger}\left[S,\Gamma\right]\phiinket.\\
	\end{split}
\end{equation}
In the second one we substitute $S^{\dagger}\Gamma=\Gamma S^{\dagger}+[S^\dagger,\Gamma]=\Gamma S^{\dagger}-[S,\Gamma]^\dagger$, where the last equality holds because $\Gamma=\Gamma^\dagger$:

\begin{equation}
	\label{eq:dgabssym2}
	\begin{split}
		\DG=&\phiinbra\Gamma-\dboxed{S^{\dagger}\Gamma} S\phiinket\\
			&=\phiinbra\Gamma-\dboxed{\left(\Gamma S^{\dagger}-[S,\Gamma]^\dagger\right)}S\phiinket\\
		&=\phiinbra\Gamma-\Gamma S^{\dagger}S+[S,\Gamma]^{\dagger}S\phiinket\\
		&=\phiinbra\Gamma\left(I-S^{\dagger} S\right)\phiinket+\phiinbra [S,\Gamma]^{\dagger}S\phiinket\\
		&=\phiinbra \Gamma A\phiinket+\phiinbra [S,\Gamma]^{\dagger}S\phiinket.\\
	\end{split}
\end{equation}

We now add the last lines of Eqs. (\ref{eq:dgabssym1}) and (\ref{eq:dgabssym2}) and divide by 2:
\begin{equation}
	\label{eq:dgabssym}
	\begin{split}
	\DG&=\phiinbra \Dabs \phiinket+\phiinbra \Dasy \phiinket\\
		&=\DG_{\mathrm{absorption}}+\DG_{\mathrm{asymmetry}}.
	\end{split}
\end{equation}

The total $\DG$ splits into two additive parts, one featuring $A$ and the other one featuring the commutator $[S,\Gamma]$. Both parts are the integrated value of a Hermitian operator, which renders them real valued and separately measurable. Additionally, when $\phiinket$ is an eigenstate of $\Gamma$, the two parts are independent of each other in the following sense. For a fixed $\phiinket$, changing the absorption without changing the commutator ($[S,\Gamma]$) affects $\DG_{\mathrm{absorption}}$ but not $\DG_{\mathrm{asymmetry}}$. Similarly, changing the commutator without changing the absorption affects $\DG_{\mathrm{asymmetry}}$ but not $\DG_{\mathrm{absorption}}$.

\subsection{Relationship with Noether's theorem}
We can connect Eq. (\ref{eq:dgabssym}) with Noether's theorem about symmetries and conserved quantities \cite{Noether1918}. We first need to particularize Eq. (\ref{eq:dgabssym}) to the case of zero absorption ($A=0$) because Noether's theorem does not apply to dissipative systems. In this case, $\DG_{\mathrm{absorption}}=0$, and Eq. (\ref{eq:dgabssym}) reduces to $\DG=\phiinbra\Dasy\phiinket$, which contains Noether's theorem in the following sense: The existence of a symmetry of the system ($[S,\Gamma]=0$) implies the existence of a conserved quantity ($\DG=0$). Notably, Eq. (\ref{eq:dgabssym}) applies to dissipative systems as well. When $A\neq 0$, the term $\phiinbra \Dabs\phiinket$ is the extra component in $\DG$ due to absorption, which is not covered by Noether's theorem.

We now aim to use the expressions obtained so far for analyzing and engineering the transfer of properties between light and matter. To such end, it is convenient to introduce basis sets for the incoming and outgoing fields.
\section{A universal formula, transfer rates, engineering bounds, and differential transfer\label{sec:basis}}
We now choose orthonormal bases sets where the eigenvalue of $\Gamma$ is one of the four defining numbers of the basis vectors\footnote{Each vector of an orthonormal basis of incoming(outgoing) transverse Maxwell fields has four numbers that identify it. These four numbers are the eigenvalues of four commuting operators. For example, multipolar fields of well-defined parity are eigenvectors of the angular momentum squared, the angular momentum along one axis, the energy (frequency), and the parity operator. Plane-waves can be chosen as eigenstates of the three components of linear momentum, which fixes the frequency, and a polarization descriptor, for example, the helicity operator. The superscripts in  $\etagammainout$ denote the incoming(outgoing) character of the fields; {\em they do not imply the relationships} $\etagammaout=S\etagammain$.}. We denote by $\etagammainout$ the basis vectors, where $\eta$ is a collective index containing the other three numbers. The incoming and outgoing states can be decomposed into such bases
\begin{equation}
	\label{eq:decomp}
		\phiinket=\sumint_{\eta,\gamma}\ain\etagammain,\
		\phioutket=\sumint_{\eta,\gamma}\aout\etagammaout,
\end{equation}
where $\ain$ and $\aout$ are complex numbers, and $\sumint_{\eta,\gamma}$ indicates that there can be summations over discrete indexes like the eigenvalues of angular momentum, or integrals over continuous variables like the eigenvalues of linear momentum or of $\mathbf{\hat{S}}$.

In this basis, the incoming and outgoing fields are represented by column vectors containing their coordinates
\begin{equation}
	\label{eq:vecs}
\ainvec\equiv\begin{bmatrix}\vdots\\\ain\\\vdots\end{bmatrix},\ \aoutvec\equiv\begin{bmatrix}\vdots\\\aout\\\vdots\end{bmatrix},
\end{equation}
and the operators $S$ and $\Gamma$ are represented by matrices whose elements are
\begin{equation}
	\label{eq:mats}
	\begin{split}
		\Smat\left(\eta\gamma,\bar{\eta}\bar{\gamma}\right)&=\baretagammaout S \etagammain,\\
		\Gammamat\left(\eta\gamma,\bar{\eta}\bar{\gamma}\right)&=\baretagammainout\Gamma\etagammainout.\ 
\end{split}
\end{equation}
for all $\left(\eta,\gamma,\bar{\eta},\bar{\gamma}\right)$. The column vectors and the matrices have infinite dimension. Since $\Gamma\etagammainout=\gamma\etagammainout$, the matrix $\Gammamat$ is diagonal in the chosen basis:
\begin{equation}
	\label{eq:gammadiag}
	\Gammamat\left(\eta\gamma,\bar{\eta}\bar{\gamma}\right)=\gamma\delta_{\gamma\bar{\gamma}}.
\end{equation}

One benefit that we obtain from this choice of basis is a formula for $\DG$ whose form is the same for any $\Gamma$.

Using Eqs. (\ref{eq:decomp}) to (\ref{eq:gammadiag}), we can write the following expressions for Eqs. (\ref{eq:transfer}) and (\ref{eq:transfer_dev}) 
\begin{eqnarray}
	\DG&=&\ainvec^{\dagger}\Gammamat\ainvec-\aoutvec^{\dagger}\Gammamat\aoutvec\\
	\label{eq:dgsm}
	&=&\ainvec^{\dagger}\left(\Gammamat-{\Smat}^\dagger \Gammamat \Smat\right)\ainvec\\
	\label{eq:dgs}
	&=&\sumint_{\eta,\gamma} \gamma\left(|\ain|^2-|\aout|^2\right).
\end{eqnarray}

Throughout the article, the appropriate differentials from $\sumint_{\eta,\gamma}$ are implicitly taken into account in the operations with vectors and matrices.

The form of Eqs. (\ref{eq:dgsm}) and (\ref{eq:dgs}) is universal in the following sense. The total transfer of a property $\Gamma$ during light-matter interactions can always be written in this form: One only needs to choose an orthonormal basis where the eigenvalue of $\Gamma$ is one of the four defining numbers. For example, when we particularize Eq. (\ref{eq:dgs}) to the $z$ component of the torque on a sphere ($\Gamma=J_z$), and use the basis of multipolar fields of well-defined parity, we recover the remarkably simple formulas that were obtained in \cite{Barton1989,Farsund1996} by different means\footnote{Since $\Gamma=J_z$, one can use the multipolar basis of well-defined $J_z$. The starting point is Eq. (\ref{eq:dgt}), which is the $\mat{T}$ matrix version of Eq. (\ref{eq:dgs}) (see App. \ref{app:ST} for more details). One then uses the fact that the $\mat{T}$ matrix of a sphere is diagonal in this basis, and its elements are the Mie coefficients. Then, the simple relationship between the coefficients of the incident and scattered fields can be used to recover the formula for the $z$ component of the torque \cite{Barton1989,Farsund1996}.}.

\subsection{Transfer rates\label{sec:transferrates}}
Given an incoming field $\ainvec$, we can use Eqs. (\ref{eq:dgsm}) or (\ref{eq:dgs}) to compute the transfer of $\Gamma$ between light and matter. We now obtain the expression for the transfer rate of $\Gamma$ per Watt of electromagnetic power in the incoming field.

We start by computing the integrated energy of the incoming field, which, in its abstract form reads: 
\begin{equation}
	E_{\textrm{in}}=\phiinbra H \phiinket,
\end{equation}
where $H$ is the energy operator, which can be represented by $i\hbar\partial_t$, where $\hbar$ is Planck's constant divided by $2\pi$.

Let us now chose a basis where frequency is well-defined. The energy in Joules of a given incoming field $\ainvec$ can be then computed as
\begin{equation}
	\label{eq:H}
	E_{\textrm{in}}={\ainvec}^{\dagger}\mat{H}\ainvec,
\end{equation}
where $\mat{H}$ is a diagonal matrix whose elements are $\hbar\omega_l$, and $\omega_l$ are the definite frequencies of the vectors of the basis. For example, if we consider a monochromatic field, then $\omega_l=\omega$ for all $l$ and the result of Eq. (\ref{eq:H}) is $\hbar\omega{\ainvec}^{\dagger}\ainvec$.

The expression
\begin{equation}
\label{eq:perjoule}
\frac{{\ainvec}^\dagger\left(\Gammamat-{\Smat}^\dagger \Gammamat \Smat\right)\ainvec}{{\ainvec}^{\dagger}\mat{H}\ainvec}
\end{equation}
is hence the transfer of $\Gamma$ between the field and the matter per Joule of energy of the incoming field. If the source of the incoming field $\ainvec$ has $W$ Joules/second (Watts) of power, the rate of transfer of $\Gamma$ $\left(d\Gamma/d t\right)$ would be $W$ times the quantity in Eq. (\ref{eq:perjoule}). For $W=1$ we have hence that 
\begin{equation}
\label{eq:transferrate}
\boxed{
\frac{d\Gamma}{d t}=\frac{{\ainvec}^\dagger\left(\Gammamat-{\Smat}^\dagger \Gammamat \Smat\right)\ainvec}{{\ainvec}^{\dagger}\mat{H}\ainvec}
}
\end{equation}
is the transfer rate per incoming Watt of electromagnetic power.

We can now use Eq. (\ref{eq:dgabssym}) to split the contributions related to absorption and asymmetry:
\begin{equation}
\label{eq:transferratesplit}
\frac{d\Gamma}{dt}=\stackrel{{\text{absorption}}}{\boxed{\frac{{\ainvec}^\dagger\left(\Dabsmat\right)\ainvec}{2{\ainvec}^{\dagger}\mat{H}\ainvec}}}+\stackrel{{\text{asymmetry}}}{\boxed{\frac{{\ainvec}^\dagger\left(\Dasymat\right)\ainvec}{2{\ainvec}^{\dagger}\mat{H}\ainvec}}}.
\end{equation}

\subsection{Engineering bounds\label{sec:engineering}}
Equations (\ref{eq:transferrate}) and (\ref{eq:transferratesplit}) are suitable for engineering purposes. For example, let us say that, given an object and a property, we seek the incoming field that maximizes the absolute transfer $\DG$ per unit power of the illumination. This can be written using Eq. (\ref{eq:transferrate}) as:
\begin{equation}
	\label{eq:max}
	\max_{\ainvec}\frac{d\Gamma}{dt}=\max_{\ainvec}\frac{{\ainvec}^\dagger\left(\Gammamat-{\Smat}^\dagger \Gammamat \Smat\right)\ainvec}{{\ainvec}^\dagger\mat{H}\ainvec}.
\end{equation}

Equation (\ref{eq:max}) is a well known optimization problem. Under the conditions of our case\footnote{Since $\Gammamat$ is Hermitian, the matrix in the numerator of Eq. (\ref{eq:max}) is Hermitian. The matrix $\mat{H}$ is Hermitian and positive definite since the frequencies are always positive: $\omega_l>0$ for all $l$.}, the maximum is equal to the largest generalized eigenvalue of $\left(\Gammamat-{\Smat}^\dagger \Gammamat \Smat\right)$ and $\mat{H}$, and is achieved when the incoming field $\ainvec_{\textrm{max}}$ is equal to a generalized eigenvector corresponding to that eigenvalue (see e.g. \cite{Cheng1965}). Similarly, the incoming field resulting in the minimum transfer rate can be found, together with such minimum rate.

More sophisticated optimization strategies are also possible. For example, we may want to maximize the transfer rate of $\Gamma$ but avoid (excessive) absorption by the system. Or we may want to maximize the transfer rate of $\Gamma_1$ without transferring (excessive) $\Gamma_2$. These kinds of constrained optimization problems can be written down using the formalism presented in this paper. The design of optimal tractor beams \cite{Chen2011,Dogariu2013} can also be conveniently achieved using the plane-wave basis. First, we restrict the elements of the vector $\vect{\alpha}$ to be non-zero only if they correspond to plane-waves with positive momentum projection along a given axis. Then, the non-zero weights are optimized for the maximum negative force along that axis.

\subsection{Differential transfer\label{sec:dtransfer}}
A particular application of light-matter interaction that is attracting research attention is chiral sorting \cite{Tkachenko2014,Cameron2014,Tkachenko2014b,Canaguier2013,Hayat2015}. Consider a mixture containing chiral objects and their mirror symmetric versions, like for example a solution of the two enantiomers of a chiral molecule. The objective is to physically separate the two groups using electromagnetic beams. In this section, we address this problem using the theoretical framework outlined so far. We assume that the direct interaction between the incoming beam and each chiral object is much stronger than the mutual interaction between objects, and do not account for the latter. Chiral sorting can be seen as differential transfer rates of linear momentum onto the two versions of the object. We perform the following analysis for a generic property $\Gamma$ since other forms of differential interaction can be considered, for example inducing different torques \cite[Sec. V]{Nieto2015b}. 

Let us consider a chiral object and its enantiomer. Their scattering operators $\Smat$ and $\Smatpar$ can be related by a transformation through the parity operator $\Pi$: $\Smatpar=\mat{\Pi}\Smat\mat{\Pi}^{-1}$. The differential transfer rate of a property $\Gamma$ given an incoming field $\ainvec$ can be obtained using Eq. (\ref{eq:dgsm}):
\begin{equation}
	\label{eq:diffrate}
	\DGS-\DGSpar={\ainvec}^\dagger\left(\Smatpar^\dagger\Gamma\Smatpar-\Smat^\dagger\Gamma\Smat\right)\ainvec.
\end{equation}
The most efficient illumination for differential transfer of $\Gamma$ is hence (Sec. \ref{sec:engineering}):
\begin{equation}
	\max_{\ainvec}\frac{{\ainvec}^\dagger\left(\Smatpar^\dagger\Gamma\Smatpar-\Smat^\dagger\Gamma\Smat\right)\ainvec}{{\ainvec}^\dagger\mat{H}\ainvec}.
\end{equation}

In situations like a solution of chiral molecules, the chiral objects are both anisotropic and freely rotating. This random orientation must be taken into account. Assuming that their rotation rates are slow enough so that the steady state conditions in Sec. \ref{sec:steadystate} apply, the right hand side of Eq. (\ref{eq:diffrate}) changes into

\begin{equation}
	\label{eq:diffrateR}
	{\ainvec}^\dagger\left[\left(\Rmatb\Smatpar\Rmatbinv\right)^\dagger\Gamma\left(\Rmatb\Smatpar\Rmatbinv\right)-\left(\Rmata\Smat\Rmatainv\right)^\dagger\Gamma\left(\Rmata\Smat\Rmatainv\right)\right]\ainvec,
\end{equation}
where $\Rmata$ and $\Rmatb$ are random rotations. The differential transfer in Eq. (\ref{eq:diffrateR}) is now a random variable: For fixed $\ainvec$, its value depends on $\Rmata$ and $\Rmatb$. We may compute its average assuming that all the orientations are equiprobable\footnote{Using obvious definitions to write Eq. (\ref{eq:diffrateR}) as ${\ainvec}^\dagger\left(\Mmatb-\Mmata\right)\ainvec$, its statistical average is $\int d\Rmata\int d\Rmatb C {\ainvec}^\dagger\left(\Mmatb-\Mmata\right)\ainvec$, where $C$ is a constant that denotes that all orientations are equiprobable. The value of $C=1$ can be deduced from the fact that $\int d\Rmata\int d\Rmatb C $ must be one, and that $\int d\Rmat=1$ (see e.g. \cite[Eq. 8.2-11]{Tung1985}). Eq. (\ref{eq:diffrateRav}) is then readily reached using that $\Mmata$($\Mmatb$) is independent of $\Rmatb$($\Rmata$), and $\ainvec$ independent of both rotations.}:

{\small
\begin{equation}
	\label{eq:diffrateRav}
	{\ainvec}^\dagger\left[\int d\Rmat \left(\Rmat\Smatpar\Rmatinv\right)^\dagger\Gammamat\left(\Rmat\Smatpar\Rmatinv\right)-\left(\Rmat\Smat\Rmatinv\right)^\dagger\Gammamat\left(\Rmat\Smat\Rmatinv\right)\right]\ainvec,
\end{equation}
}
where each rotation $\Rmat$ is characterized by its three Euler angles $[\phi,\theta,\psi]$, and $\int d\Rmat=\frac{1}{8\Pi^2}\int_0^\Pi d\theta\sin\theta\int_0^{2\pi}d\phi\int_0^{2\pi}d\psi$.

The integral in Eq. (\ref{eq:diffrateRav}) defines a Hermitian matrix $\mat{M}$. Therefore, the solution of the optimization problem
\begin{equation}
	\label{eq:soldiff}
\max_{\ainvec}\frac{{\ainvec}^\dagger\mat{M}\ainvec}{{\ainvec}^\dagger\mat{H}\ainvec}
\end{equation}
produces both the most efficient incoming beam for the rotationally averaged differential transfer of $\Gamma$ between the two enantiomeric versions of the object, and the actual value of such upper bound.

\section{Practical implementation\label{sec:implementation}}
In this section we develop the formalism further into a computer friendly formulation. We address the questions of how to practically obtain the $\Smat$ matrix of a given object, and how to deal with its theoretically infinite dimension. Both questions are resolved by the T-matrix.

There exists a convenient practical way of obtaining the scattering matrix $\Smat$ for the implementation of the proposed methodology. The scattering matrix $\Smat$ can be numerically obtained as:
\begin{equation}
	\label{eq:smattmat}
\Smat=\underline{\underline{I}}+2\Tmat,
\end{equation}
where $\mat{I}$ is the identity matrix and $\Tmat$ is the T-matrix, a.k.a transfer matrix. There is abundant literature on the T-matrix. Reference \onlinecite{Mishchenko2016} is a comprehensive collection of T-matrix references classified by themes. 

There are also publicly available computer codes for computing the T-matrix for different kinds of objects \cite{Mishchenko2000,Nieminen2007}. These numerical tools produce finite T-matrices in the multipolar basis of well-defined parity. The dimensions of the matrix are set by the selection of a maximum multipolar order, which must be large enough so that the interaction of the object with the multipolar fields corresponding to the discarded higher orders can be neglected according to some criteria. 

\begin{figure}[h!]
	\begin{center}
	\includegraphics[width=\linewidth]{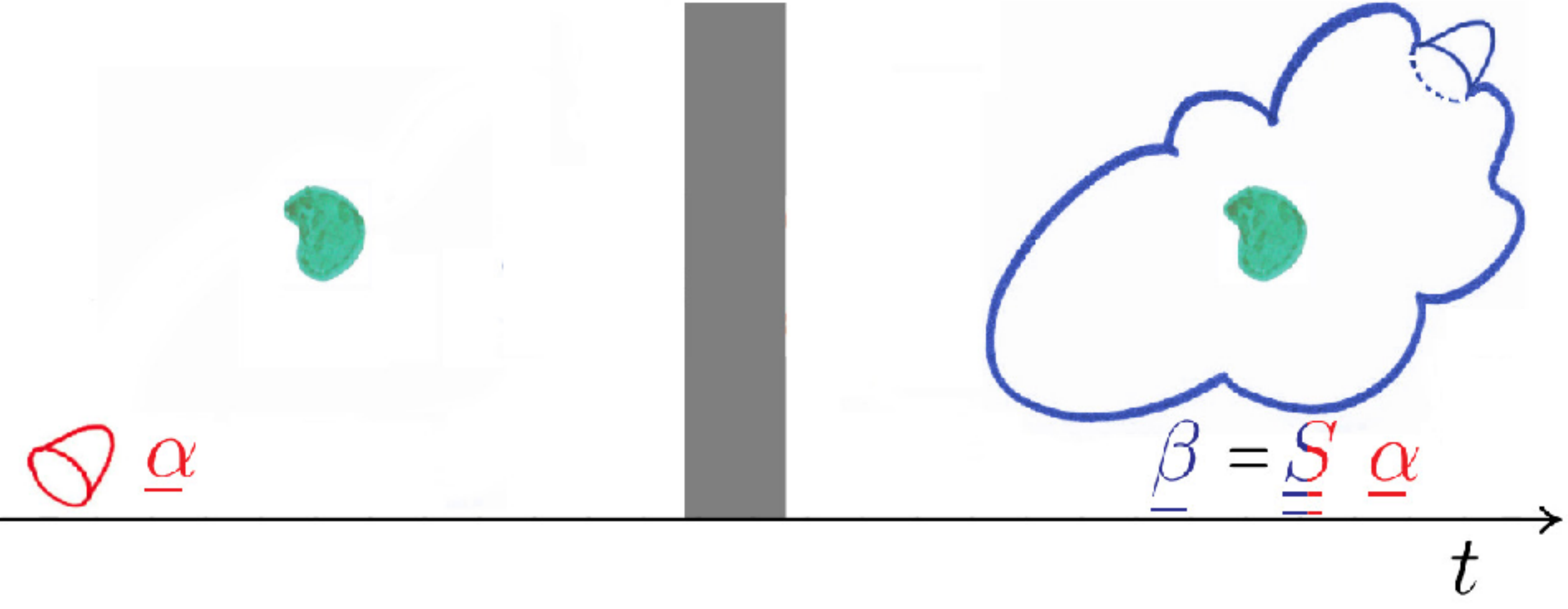}
		\end{center}
\caption{\label{fig:smatrix} After choosing basis sets for the incoming and outgoing fields of Fig. \ref{fig:scatt_op}, the vector of coordinates of the incoming field $\vect{\alpha}$, and that of the outgoing field $\vect{\beta}$ are related by the matrix representation of the scattering operator: $\vect{\beta}=\mat{S}\vect{\alpha}$. In this setting, the $\Smat$ matrix connects purely incoming fields to purely outgoing fields.}
\end{figure}

\begin{figure}[h!]
	\begin{center}
		\includegraphics[width=0.75\linewidth]{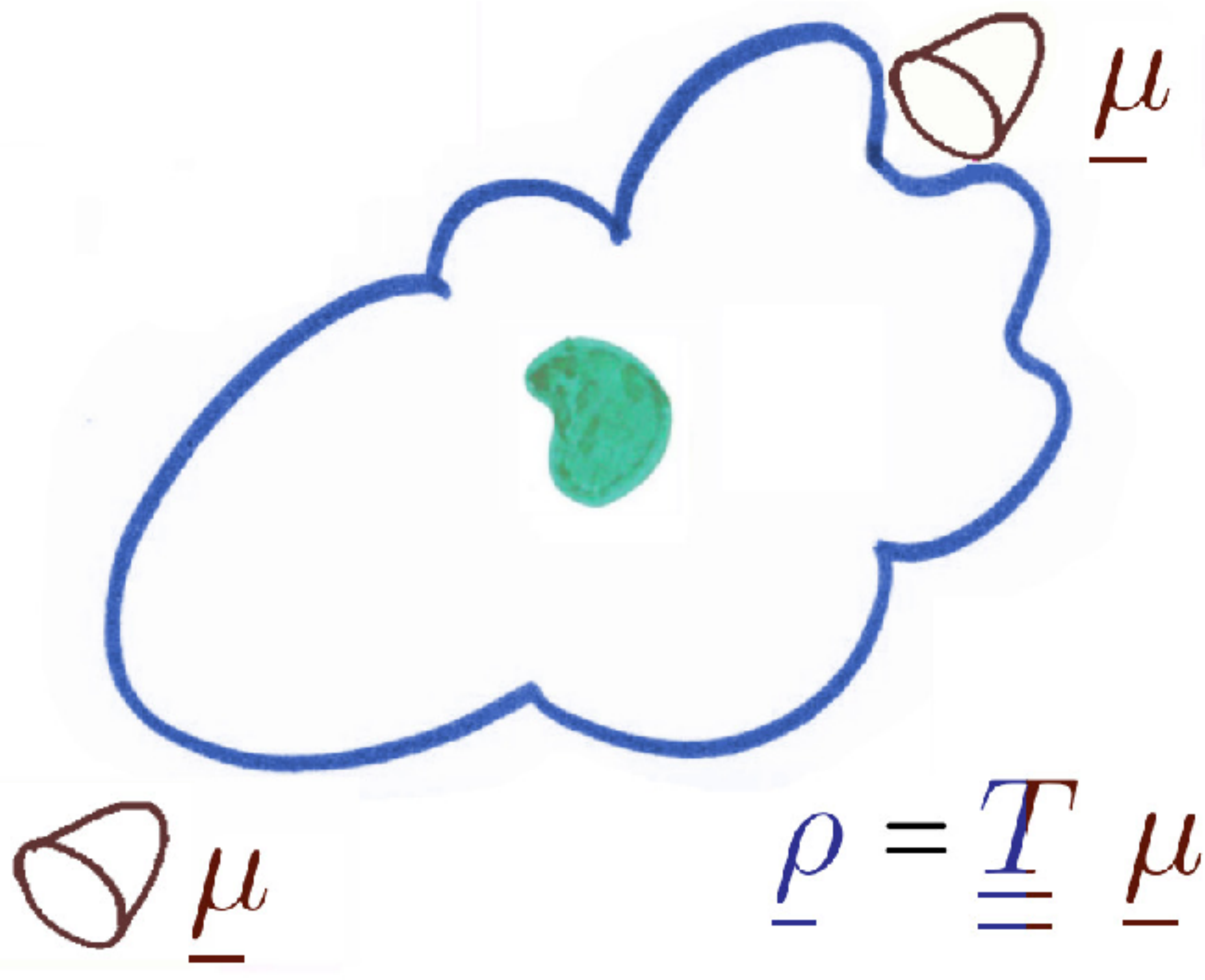}
						\end{center}
\caption{\label{fig:tmatrix} In the T-matrix setting, the matrix $\mat{T}$ relates incident fields represented by the coordinate vector $\vect{\mu}$ to scattered fields represented by  $\vect{\rho}$: $\vect{\rho}=\mat{T}\vect{\mu}$. In this setting, the incident fields have a {\em mixed} incoming and outgoing character, and the scattered field is equal to the total outgoing field ($\vect{\beta}$ in Fig. \ref{fig:smatrix}) {\em minus} the outgoing part of the incident field (equal to $\vect{\mu}/2$). As explained in the text, the two settings in Figs. \ref{fig:smatrix} and \ref{fig:tmatrix} have a tight numerical relationship, but exhibit important differences in their physical interpretation.}
\end{figure}

Notwithstanding the simple numerical relationship in Eq. (\ref{eq:smattmat}), there is an important difference between the physical meaning of the two matrices. The $\Smat$ matrix maps incoming fields $\ainvec$ to outgoing fields $\aoutvec$: $\aoutvec=\Smat\ainvec$. The $\Tmat$ matrix maps incident fields $\incvec$ to scattered fields $\scattvec$: $\scattvec=\mat{T}\incvec$. We depict the two different settings in  Figs. \ref{fig:smatrix} and \ref{fig:tmatrix}, respectively. While the scattered field $\vect{\rho}$ is only outgoing, the incident field $\vect{\mu}$ contains both incoming and outgoing parts. This is a physically relevant difference. For example, while measuring the complete scattered field requires interferometric techniques to cancel the outgoing part of the incident field from the total outgoing field \cite{Husnik2012}, the measurement of incoming and/or outgoing fields does not. 

The $\Smat$ matrix setting is the one which is directly usable for our purpose of subtracting the incoming and outgoing fluxes of some property in order to compute its transfer from the field to the object. Nevertheless, Eq. (\ref{eq:smattmat}) allows to benefit from existing T-matrix codes. Appendix \ref{app:ST} contains the numerical relationships between the objects of the two settings, as well as the expressions of $\DG$ in the T-matrix setting.
We now illustrate the proposed methodology with two examples. 

\section{Examples\label{sec:examples}}
For the first example, we choose a rather complex system in order to showcase the practical capabilities of the approach. In particular, that our methodology is exact, and can be used for any object, including those whose sizes place them outside the ranges where the dipolar approximation or the ray optics approximation are suitable. The object that we choose is chiral and, effectively, spherically symmetric. This kind of objects are often considered in the literature \cite{Moloney2007,Tkachenko2014b,Canaguier2015,Hayat2015,Fernandes2016}.

For the second example, we choose an electromagnetically small uniaxial chiral object, with which we illustrate the chiral sorting theory of Sec. \ref{sec:dtransfer}.

\begin{table}[]
	\label{tab:mat}
	\begin{ruledtabular}
\begin{center}\begin{tabular}{ccc|cc}
	& {$|\omega \ j\ m\ \lambda\rangle^{\textrm{in(out)}}$}  & & {\hspace{0.5cm}$|\omega \ \theta\ \phi\ \lambda\rangle^{\textrm{in(out)}}$} & \\
	\hline
	$H$ & $\Lambda$ & $J_z$ & $P_z$ & $\hat{S}_z=\Lambda P_z/|\mathbf{P}|$\\
	$\hbar\omega_l$ & $\hbar\lambda_l$ & $\hbar m_l$ & $\hbar \frac{\omega}{c}\cos\theta_l$ & $\hbar\cos\theta_l$
\end{tabular}\end{center}
	\end{ruledtabular}
	\caption{\label{tab:gammamatrices} Expression of the matrix representation of different operators: Energy $H$, helicity $\Lambda$, and the third components of angular momentum $J_z$, linear momentum $P_z$, and ``spin angular momentum'' $\hat{S}_z$. The matrices are diagonal in appropriate bases. For $H$, $\Lambda$ and $J_z$ we choose the basis of multipoles of well-defined helicity $|\omega \ j\ m\ \lambda\rangle$. For $P_z$ and $\hat{S}_z$, the basis of plane-waves of well-defined helicity $|\omega \ \theta\ \phi\ \lambda\rangle$, where the momentum of the plane-wave is $\mathbf{p}=\frac{\omega}{c}[\sin\theta\cos\phi,\sin\theta\sin\phi,\cos\theta]$. The last row of the table shows the diagonal elements of the matrices representing each operator in their corresponding bases. The matrices are the same for both incoming and outgoing versions of the bases. See App. \ref{app:num} for more details.}
\end{table}

\subsection{Example 1\label{sec:example1}}
\begin{figure*}[p]
		\subfloat[]{
					\begin{overpic}[width=0.46\textwidth]{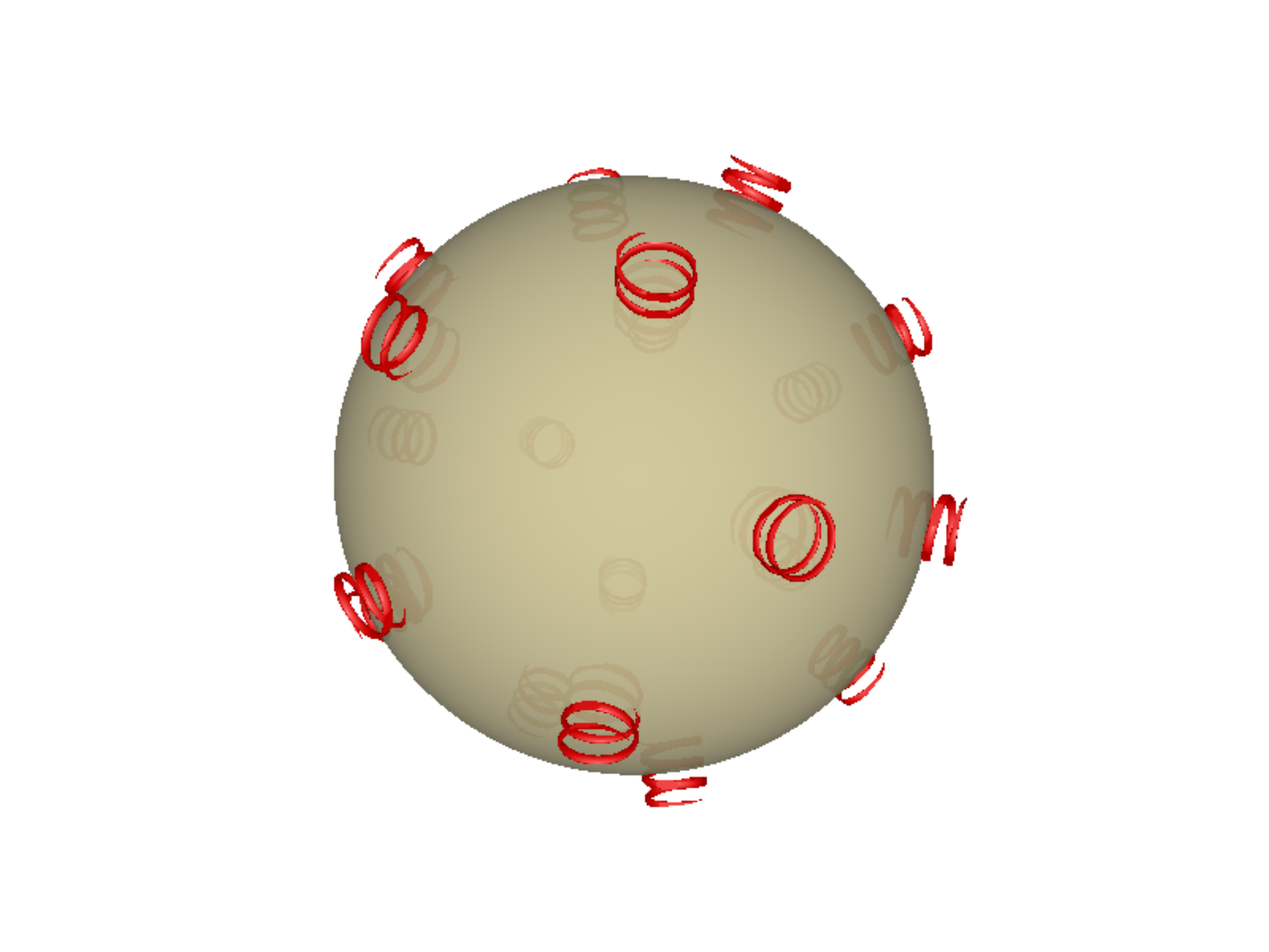}
			\put(20,60){{\small (a)}}
\end{overpic}}
		\subfloat[]{\begin{overpic}[width=0.46\textwidth]{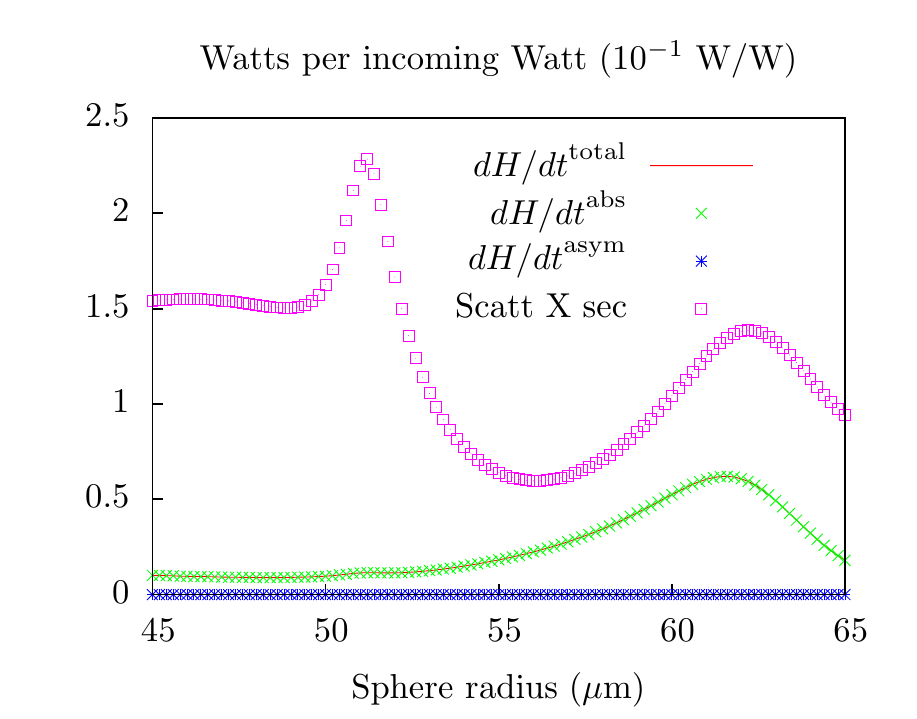}
		\put(25,60){(b)}
\end{overpic}}\\
		\subfloat[]{\begin{overpic}[width=0.46\textwidth]{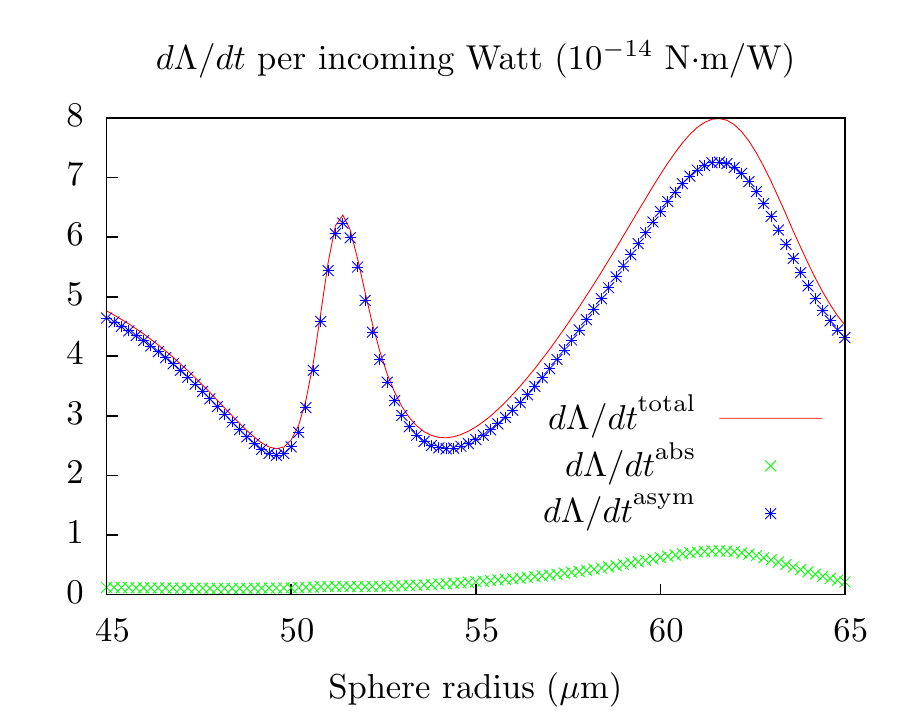}
		\put(25,60){(c)}
\end{overpic}}
		\subfloat[]{\begin{overpic}[width=0.46\textwidth]{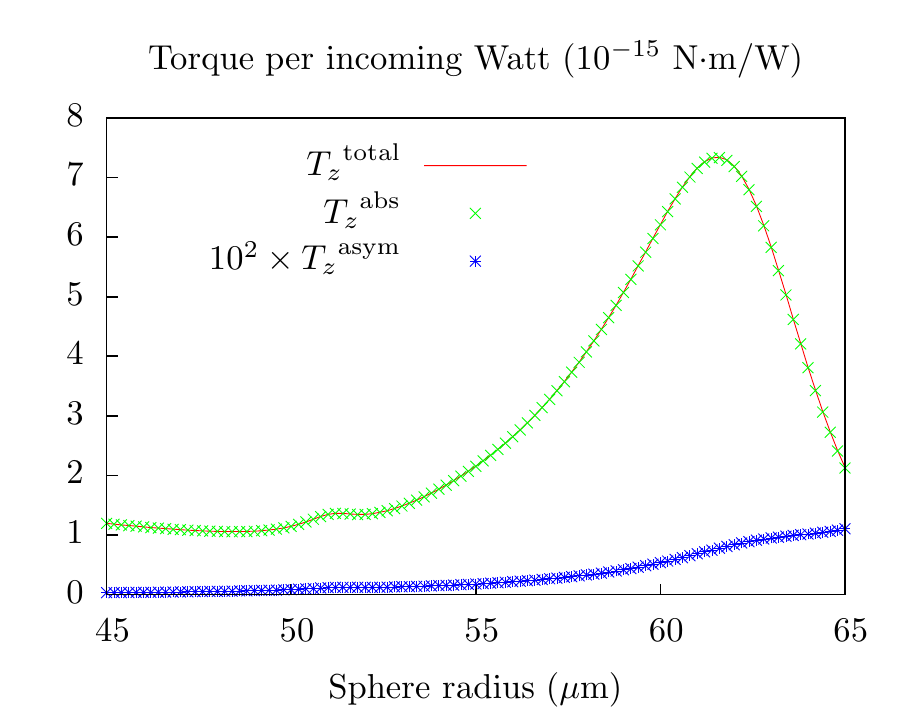}
		\put(25,60){(d)}
\end{overpic}}\\
	\subfloat[]{\begin{overpic}[width=0.46\textwidth]{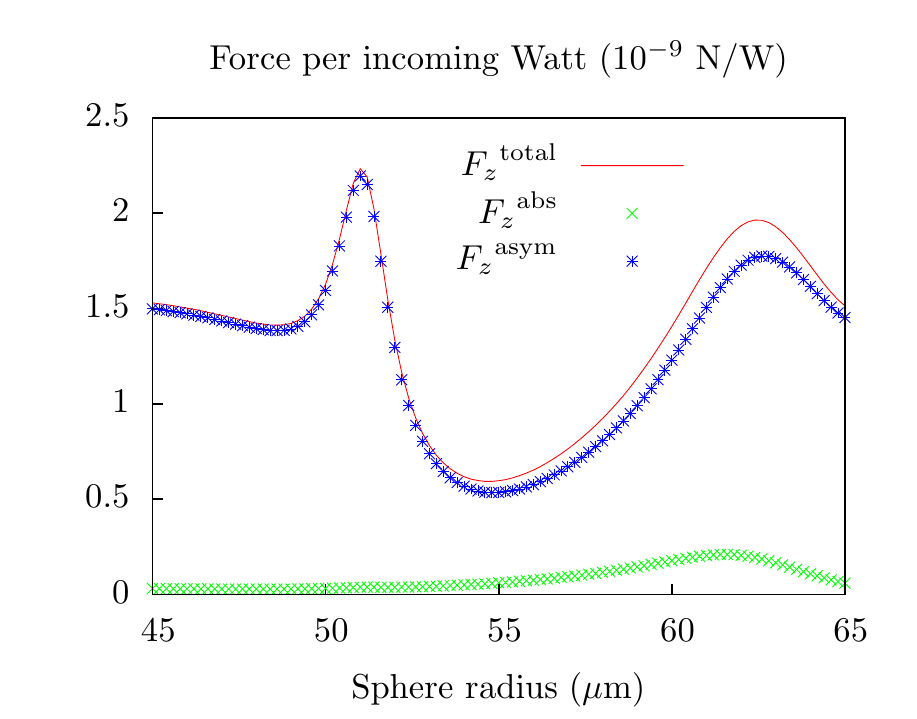}
		\put(25,60){(e)}
\end{overpic}}
	\subfloat[]{\begin{overpic}[width=0.46\textwidth]{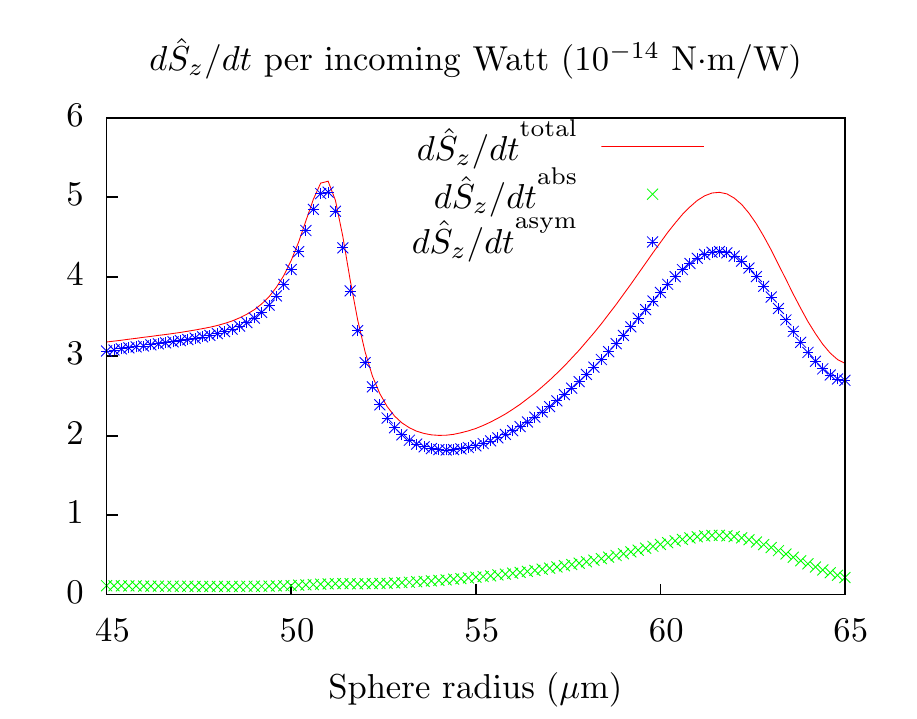}
		\put(25,60){(f)}
	\end{overpic}}
\caption{\label{fig:rates}The object in panel (a) is illuminated by a single monochromatic plane-wave (not shown). The material system is composed of a CdSe sphere decorated with twenty helical objects in an icosahedral arrangement. The plane-wave is of helicity $+1$ (i.e. left handed polarized), vacuum wavelength equal to $222.8$ $\mu$m, and momentum aligned with the positive $z$ direction. Panels (b) to (f) show the transfer rates of different properties between the field and the object as a function of the CdSe sphere radius: Energy (b), helicity (c), $z$-component of the angular momentum (torque) (d), $z$-component of the linear momentum (force) (e), and $z$-component of the transverse ``spin angular momentum'' (f). All quantities are expressed in SI units. The plots contain the total transfer rates (red lines), the transfer rates due to absorption (green crosses), and the transfer rates due to asymmetry (blue stars). Panel (b) additionally contains the scattering cross-section that the object presents to the illumination (magenta squares).}
\end{figure*}

We consider a material system composed of a CdSe sphere decorated with twenty helical objects in an icosahedral arrangement [see Fig. \ref{fig:rates}(a)]. The composite object is surrounded by vacuum. This is an example of a chiral object which is practically symmetric under rotations. Spherically symmetric chiral objects are often theoretically and experimentally studied \cite{Moloney2007,Tkachenko2014b,Canaguier2015,Hayat2015,Fernandes2016}. We highlight that the method by which we obtain the electromagnetic response of our object is free of approximations like the dipole approximation, ray optics, or chiral constitutive material parameters $\chi\neq0$.

While the expressions obtained in the previous sections allow for wide-band poly-chromatic illuminations, we fix here the operating frequency in order to simplify the example. We will take the monochromatic field as an approximation for a narrowband beam, and describe the object by the $\Smat$ matrix at the central frequency. This assumes that the response of the object does not substantially change within the bandwidth of the beam. We have chosen the frequency so that this is met. 

From another point of view, the case of a wide-band poly-chromatic illumination on a time-invariant system\footnote{Time invariance ensures that the interaction does not couple different frequencies.} can be treated by an integral over many monochromatic illuminations with a frequency dependent $\Smat$. The calculations for each frequency would then be similar to those presented in this section.

We will also assume a time invariant $\Smat$. This can be seen as one of the individual steps in the step wise steady state procedure discussed in Sec. \ref{sec:steadystate}. 

We will first analyze the transfer rate of several quantities between the material system and a single plane-wave as a function of the radius of the CdSe sphere. The separation between absorption and asymmetry mediated transfers will be considered. Then, we will compare the force and helicity transfer rate per incoming Watt exerted by the single plane-wave to their corresponding upper bounds. 

The calculations are based on Eqs. (\ref{eq:transferrate}) and (\ref{eq:transferratesplit}). Appendix \ref{app:num} contains the detailed explanation of how to obtain all the quantities that appear in these equations, plus a few extra practical implementation notes including indications on how to compute transfer rates of components of $\mathbf{P}$, $\mathbf{J}$, and $\mathbf{\hat{S}}$, in arbitrary directions. 

Figure (\ref{fig:rates}) shows the transfer rates per incoming Watt of several quantities as a function of the radius of the CdSe sphere. The quantities are: Energy $H$ [Fig. \ref{fig:rates}(b)], helicity $\Lambda$ [Fig. \ref{fig:rates}(c)], and the third components\footnote{The transfer of the other Cartesian components is comparatively small in our example.} of angular momentum $J_z$ [Fig. \ref{fig:rates}(d)] (torque), linear momentum $P_z$ [Fig. \ref{fig:rates}(e)] (force), and ``spin angular momentum'' $\hat{S}_z$ [Fig. \ref{fig:rates}(f)] (\cite{FerCor2013b}). All quantities are expressed in SI units. Table \ref{tab:gammamatrices} contains the matrix representation of the operators. We have used Eq. (\ref{eq:transferratesplit}) to separately show the contributions due to absorption and asymmetry.

\afterpage{\clearpage}
	\vspace{-0.25cm}

Figure \ref{fig:rates}(b) also contains the scattering cross-section that the object presents to the illumination. We see in Fig. \ref{fig:rates}(b) that all the energy transfer is due to absorption. This must be the case because the system is time invariant: It has the time translation symmetry generated by the energy operator. Similarly, the approximate spherical symmetry of the system forces the torque $T_z=dJ_z/dt$ in Fig. \ref{fig:rates}(d) to be mostly due to absorption as well, hence the coincidence in shape between $dH/dt$ and $T_z$ [Figs. \ref{fig:rates}(b) and (d)]. On the other hand, Fig. \ref{fig:rates}(c) shows that helicity is mostly transferred due to the broken duality symmetry of the object (\cite{FerCor2012p}, \cite[Sec. 2.7]{FerCorThesis}), and absorption plays a lesser role. Figures \ref{fig:rates}(e) and (f) reveal that both $F_z=dP_z/dt$ and $d\hat{S}_z/dt$ are also mostly due to asymmetry. The object breaks translational symmetry along the $z$ axis, and is susceptible to be pushed by the plane-wave in this direction. The object is also far from being symmetric under the transformations generated by $\hat{S}_z=\Lambda\frac{P_z}{|\mathbf{P}|}$, which are helicity and frequency dependent {\em translations}, as can be foreseen by the expression for $\mathbf{\hat{S}}=\Lambda\mathbf{P}/\mathbf{|P|}$, where $\Lambda$ is the helicity operator and $\mathbf{P}$ the momentum vector operator \cite{FerCor2013b}. An example of an object with the symmetry generated by $\hat{S}_z$ is an infinitely long (in $z$) dual symmetric object with constant cross-section in the XY plane.

The positions of the common feature of helicity, momentum and $\hat{S}_z$ transfer in Figs. \ref{fig:rates}(c), (e), and (f), correlate well with the maximum in the scattering cross-section in Fig. \ref{fig:rates}(a).

\begin{figure}[h]
		\subfloat[]{\includegraphics[width=\linewidth]{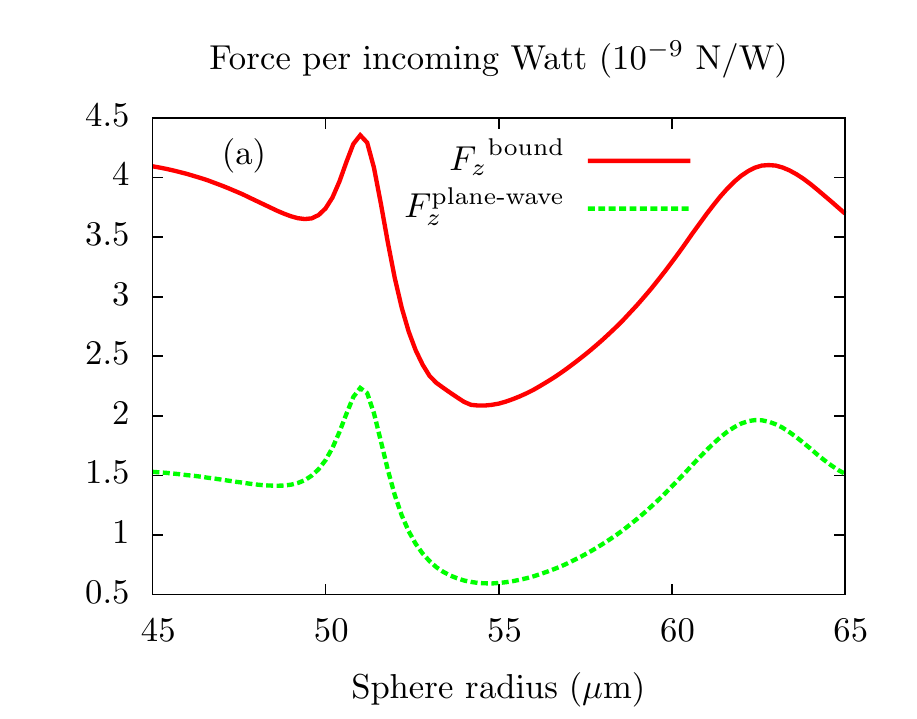}}\\
		\subfloat[]{\includegraphics[width=\linewidth]{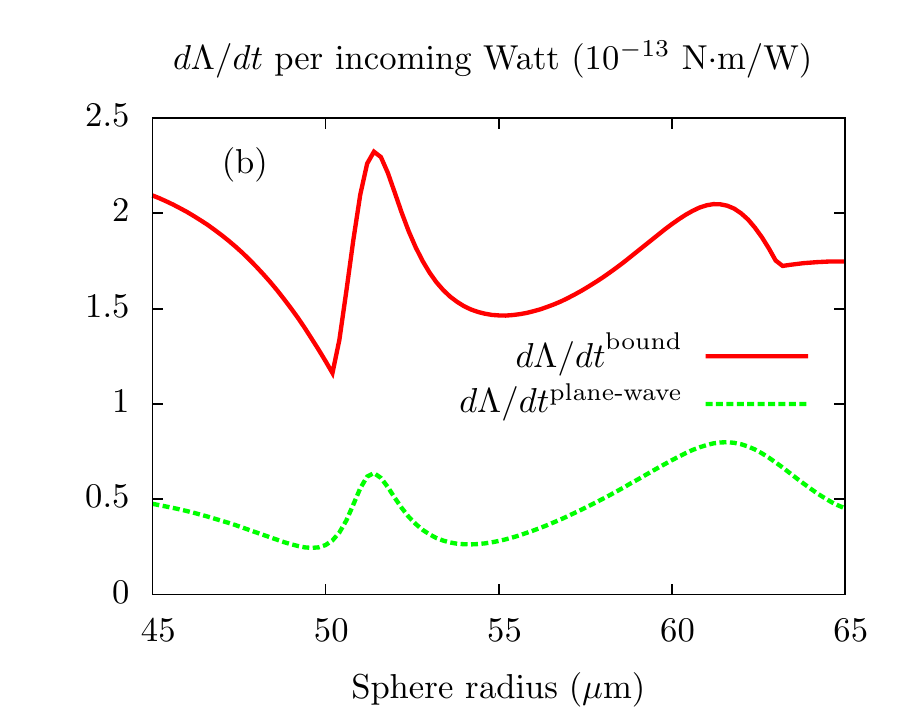}}\\
		\caption{\label{fig:max_f3}The green dashed lines show the force per incoming Watt of power (a) and the helicity transfer (b) for the plane-wave used in Fig. \ref{fig:rates}. The red solid lines show the upper bounds of the same quantities set by the optimal monochromatic illuminations of the same frequency for each value of the sphere radius.}
\end{figure}

Finally, Fig. \ref{fig:max_f3} shows the force $F_z$ and helicity transfer $d\Lambda/dt$ due to the plane-wave compared to the achievable maxima for the optimal monochromatic illuminations for each sphere radius. The maxima are obtained as indicated in Sec. \ref{sec:engineering}: They are the largest positive eigenvalue of the generalized eigenvalue decomposition of the matrices $\Gammamat-\Smat^{\dagger}\Gammamat\Smat$ and $\mat{H}$. For the force $\Gammamat=\mat{P_z}$, and for the helicity transfer $\Gammamat=\mat{\Lambda}$.

\subsection{Example 2\label{sec:example2}}

\begin{figure}[h]
		\includegraphics[width=\linewidth]{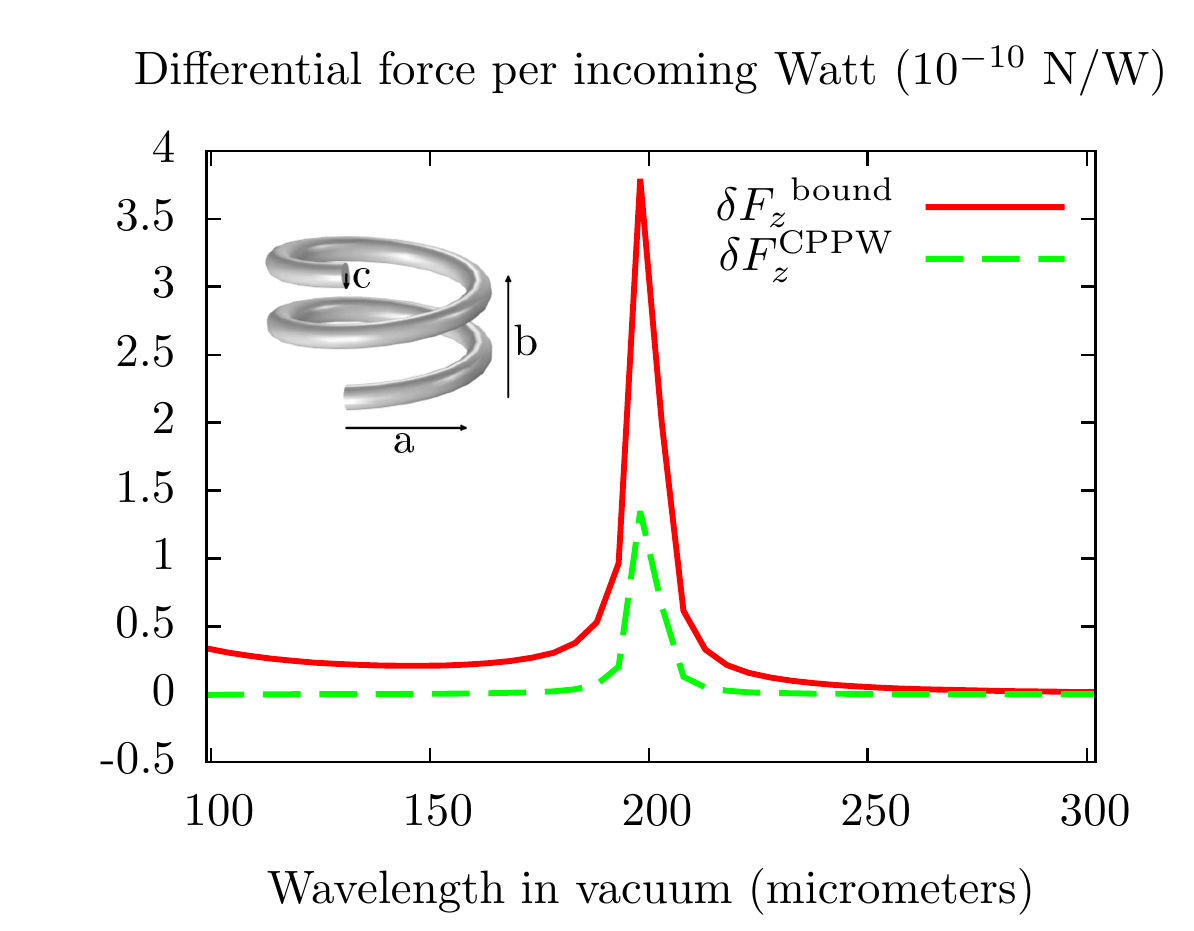}
		\caption{\label{fig:diff}Difference between the rotationally averaged force exerted onto two chiral objects by two beams: The optimal monochromatic beam for each frequency ($\delta{F_z}^{\textrm{bound}}$), and a beam composed by the sum of two monochromatic counter-propagating plane-waves of opposite helicity aligned with the $z$ axis ($\delta F_z^{\textrm{CPPW}}$). The chiral objects are the two enantiomeric versions of the silver helix shown in the inset. The dimensions of the helix are: Major radius $a=6.48$ $\mu$m, height $b=8.52$ $\mu$m, and wire radius $c=0.8$ $\mu$m. }
\end{figure}

This example is an application of the differential transfer theory of Sec. \ref{sec:dtransfer}. We will compute the differential force between the two enantiomeric versions of a chiral object: The silver helix in the inset of Fig. \ref{fig:diff}. In the considered frequency range, the helix is practically a dipolar object: Its dipolar response is at least 500 times larger than all the other higher orders added together. We have studied this object in \cite[Fig. 5]{FerCor2016}. The elements of its polarizability tensor can be seen in \cite[Fig. 3]{Rahimzadegan2016b}. The T-matrix of the helix was obtained by rigorous full-wave simulations \cite{Fruhnert2016b}.

For each frequency, we compute the rotationally averaged differential force achieved in the $z$ direction by two monochromatic illuminations. One is the optimal beam as defined by Eq. (\ref{eq:diffrateRav}). The other is composed by the sum of two counter-propagating plane-waves of opposite helicity aligned with the $z$ axis \cite{Fernandes2016}.  Figure \ref{fig:diff} shows the performance of the two beams. In both cases, the force has its maximum at the same frequency. This is related with the facts that the helix approaches the upper bound of electromagnetic chirality around that frequency (see \cite[Fig. 5]{FerCor2016}), and that objects that achieve such upper bound are transparent to one of the helicities of the field.

\section{Concluding remarks\label{sec:conclusions}}
We have presented a unified theory to treat conservation laws in light-matter interactions. It can be used to describe and engineer the transfer of any measurable property from the electromagnetic field to any object. The theory allows to explicitly characterize and separately compute the transfer due to asymmetry of the object and the transfer due to field absorption by the object. The two components are separately measurable. Among other general engineering possibilities, the framework also allows to compute the upper bound of the transfer rate of any given property to any given object, together with the corresponding most efficient illumination which achieves the bound. The approach is inherently suited for computer implementation. 

These features make us believe that the approach will be useful in the expanding field of optical manipulation, in particular in chiral light-matter interactions where conservation laws like the one for helicity are envisioned to play an important role.

\begin{acknowledgments}
I.F-C wishes to warmly thank Ms. Magda Felo for her help with the figures. We also thank Dr. Martin Fruhnert for providing us with the T-matrix of the silver helix.
\end{acknowledgments}

\appendix
\section{Singular value decomposition of the $S$ and $A$ operators\label{app:svd}}
The singular value decomposition of any complex matrix $\mat{X}$ always exists, meaning that $\mat{X}$ can always be written as:
\begin{equation}
	\mat{X}=\mat{W}\mat{D}\mat{V}^{\dagger},
\end{equation}
where $\mat{W}$ and $\mat{V}$ are unitary matrices and $\mat{D}$ is a diagonal matrix containing real numbers $d_l\ge 0$ and $d_1\ge d_2 \ge d_3 \ldots$. 

The singular value decomposition exists also for completely continuous operators \cite[Chap. II, \S 2]{Gohberg1969}. The scattering operator $S$ is completely continuous for interactions with finite cross-sections \cite[Chap. 8.6]{Weinberg2013}, which renders $A=I-S^{\dagger}S$ completely continuous as well.

As seen in Sec. \ref{sec:basis}, completely continuous linear operators can be represented by complex matrices of infinite dimension. We adopt such notation here.

Let us now consider the singular value decomposition of $\Smat$:
\begin{equation}
	\label{eq:Ssvd}
	\mat{S}=\mat{W}\mat{D}\mat{V}^{\dagger}=\sum_{l} d_l\vect{w}_l{\vect{v}_l}^{\dagger}.
\end{equation}
We may say that the $\mat{W}$ and $\mat{V}$ basis are the ``probe'' and ``measurement'' basis adapted to the object, in the following sense: When the incoming field is $\vect{v}_l$, the outgoing field is all in a single mode $d_l\vect{w}_l$. Furthermore, the scattering coefficient $\vect{w}_l^\dagger\mat{S}\vect{v}_l=d_l$ is guaranteed to be real and positive. This means that a single experimental measurement of its power $|\vect{w}_l^\dagger\mat{S}\vect{v}_l|^2$ is enough to determine it. For a general incoming field and measurement mode, the scattering coefficient is complex, and the determination of its phase is a non-trivial task. 

Let us now analyze the relationship between the singular value decompositions of the scattering and absorption operator, $\Smat$ and $\mat{A}=\mat{I}-\Smat^\dagger \Smat$ [Eq. (\ref{eq:A})], respectively. Given the decomposition of $\mat{S}$ in Eq. (\ref{eq:Ssvd}), it is straightforward to show that
\begin{equation}
	\label{eq:Asvd}
	\mat{A}=\sum_{l} (1-d_l^2)\vect{v}_l{\vect{v}_l}^{\dagger}.
\end{equation}
Equation (\ref{eq:Asvd}) is reached by first substituting the singular value decomposition of $\Smat$ (Eq. \ref{eq:Ssvd}) into the definition of $\mat{A}=\mat{I}-\Smat^\dagger \Smat$: We get $\mat{A}=\mat{I}-\mat{V}\mat{D}^\dagger\mat{W}^{\dagger}\mat{W}\mat{D}\mat{V}^{\dagger}$. The, since $\mat{W}^{\dagger}\mat{W}=\mat{I}$ and $\mat{I}=\mat{V}\mat{V}^{\dagger}$, we reach $\mat{A}=\mat{V}\left(\mat{I}-\mat{D}^2\right)\mat{V}^{\dagger}$, which is a different way of writing Eq. (\ref{eq:Asvd}).

A salient fact is that the $\vect{v}_l$'s($\vect{w}_l$'s) form a basis for the incoming(outgoing) fields. Then, we can say that the energy of a general incoming field contained in each right singular vector $\vect{v}_l$ is split by the object into two parts: One is absorbed by the object, the other goes to the outgoing field, but only into a single outgoing mode: The corresponding left singular vector $\vect{w}_l$. The ratio between re-radiated and absorbed power is $(d_l)^2/(1-d_l^2)$. Here, $d_l$ is the corresponding singular value. Note that no gain implies $d_l\le 1$, and neither gain nor loss implies $d_l=1$. Along these lines, we can also interpret $(1-d_l^2)/(d_l)^2$ as the fraction of energy that changes its character from propagating to localized.

This analysis helps making physical considerations. For example, regarding the design of perfect absorbers \cite{Alaee2012,Grigoriev2015}: The object can act as a perfect absorber for some incoming field if and only if there exist singular values equal to zero. In such case, the most general field that is completely absorbed by the object is a linear combination of the $\vect{v}_l$'s with zero singular value. 

\section{Relations between the $\Smat$ and $\Tmat$ settings\label{app:ST}}
The tight numerical relation between the two approaches depicted in Figs. \ref{fig:smatrix} and \ref{fig:tmatrix} is established using multipolar fields as basis vectors, and equating the incoming and outgoing parts in each approach. In the T-matrix approach, the input vector is expanded in regular multipoles, which contain the spherical Bessel functions $j_l(\cdot)$, and the output vector is expanded in outgoing multipoles, which contain the spherical Hankel functions of the first kind $h^+_l(\cdot)$. In the $\Smat$ matrix, the input vector is expanded in incoming multipoles, which contain the spherical Hankel functions of the second kind $h^-_l(\cdot)$, and the output vector is expanded in outgoing multipoles. The relationships between incoming/outgoing and regular spherical functions is:
\begin{equation}
	\label{eq:sph}
	\begin{split}
		h^+_l(\cdot)&=j_l(\cdot)+in_l(\cdot),\\
		h^-_l(\cdot)&=j_l(\cdot)-in_l(\cdot),\\
	\end{split}
\end{equation}
where $n_l(\cdot)$ is the spherical Neumann function.

The numerical relationship between the two approaches is obtained by equating the incoming and outgoing parts. It follows from Eq. (\ref{eq:sph}) that $j_l(\cdot)=\left[h^+_l(\cdot)+h^-_l(\cdot)\right]/2$. Physically, this means that the input vector $\vect{\mu}$ in the T-matrix setting contains both incoming and outgoing parts. Since the input vector $\vect{\alpha}$ in the $\Smat$ matrix only contains incoming parts, it follows that:

\begin{equation}
	\label{eq:inc}
\ainvec=\incvec/2.
\end{equation}
For the outgoing part, we equate the output vector $\vect{\beta}$ of the $\Smat$ matrix case to the sum of the scattered vector $\vect{\rho}$ and the outgoing part of the input vector of the T-matrix case
\begin{equation}
	\label{eq:out}
\aoutvec=\scattvec+\incvec/2.
\end{equation}

We now reach the key relationship
\begin{equation}
	\label{eq:b4}
	\begin{split}
		\aoutvec=\Smat\ainvec&
		\impliesdueto{\text{Eqs. (\ref{eq:inc})-(\ref{eq:out})}}\scattvec+\incvec/2=\Smat\incvec/2\\
		&\impliesdueto{\vect{\rho}=\mat{T}\vect{\mu}} \Tmat\incvec+\incvec/2=\Smat\incvec/2\\
		&\implies \left(\mat{I}+2\Tmat\right)\incvec=\Smat\incvec.\\
	\end{split}
\end{equation}
Since Eq. (\ref{eq:b4}) is met for any $\incvec$, it implies Eq. (\ref{eq:smattmat}): $\Smat=\mat{I}+2\Tmat$.

To sum up, the numerical relationship between the quantities in both settings is 
\begin{equation}
	\label{eq:tab}
	\begin{split}
		\Smat&=\mat{I}+2\Tmat,\\
		\scattvec=\Tmat\incvec&,\ \aoutvec=\Smat\ainvec,\\
		\ainvec=\incvec/2 &,\ \aoutvec=\scattvec+\incvec/2.
	\end{split}
\end{equation}

The expressions corresponding to Eqs. (\ref{eq:dgs}) and (\ref{eq:dgsm}) in the T-matrix language are reached using Eq. (\ref{eq:tab}):

\begin{equation}
	\label{eq:dgt}
	\begin{split}
		\DG=&\sumint_{\eta,\gamma} \gamma\left(|\ain|^2-|\ain+\scatt|^2\right)=\\
			&\sumint_{\eta,\gamma} \gamma\left(2\mathbb{R}\left\{{\ain}^*\scatt\right\}-|\scatt|^2\right)=\\
			&\sumint_{\eta,\gamma} \gamma\left(\mathbb{R}\left\{{\inc}^*\scatt\right\}-|\scatt|^2\right).
	\end{split}
\end{equation}

\begin{equation}
	\label{eq:dgtm}
	\begin{split}
		\DG=&-\frac{1}{2}{\incvec}^\dagger\left(\Gammamat\Tmat+{\Tmat}^\dagger\Gammamat+2{\Tmat}^\dagger\Gammamat\Tmat\right){\incvec}\\
		&=-{\incvec}^\dagger\left(\mathbb{R}\left\{\Gammamat\Tmat\right\}+{\Tmat}^\dagger\Gammamat\Tmat\right){\incvec},
	\end{split}
\end{equation}
where the last equality holds because $\Gamma$ is a Hermitian operator.

\section{Practical implementation notes\label{app:num}}
\subsection{The calculations for Figs. \ref{fig:rates} and \ref{fig:max_f3}}
The main ingredient in the calculations is the $\Smat$ matrix of the composite object. The $\Smat$ matrix is obtained using its $\mat{T}$ matrix and Eq. (\ref{eq:smattmat}). The T-matrix of the composite object can be obtained from the individual response of the sphere and that of a single helix. The $\mat{T}$ matrix of the sphere is obtained analytically from its Mie coefficients. The permittivity of CdSe is taken from \cite{Lisitsa1969}. The T-matrix of an individual helix in Fig. \ref{fig:rates}(a) can be seen as a scaled up model of that of a uniaxial chiral molecule. It features a dominant electric polarizability, whose amplitude is about $100$ times larger than the cross-polarizability terms, and $10^5$ times larger than the magnetic polarizability. The total $\mat{T}$ matrix of the system is obtained by rigorously solving the mutually coupled system of the sphere and the twenty rotated and translated helices [see Fig. \ref{fig:rates} (a)]. The maximum multipolar order that we consider is $j=4$ both for the mutual interaction between each element, and for the final T-matrix of the system. The calculations show that the total contributions of the neglected higher orders with $j>4$ can at most represent a portion equal to $5e\text{-}5$ of the total Frobenius norm squared of the exact (infinite dimensional) version of the T-matrix.

Another ingredient in the calculation is the expression of the $\Gamma$ matrix for the different properties that we analyze: Energy $H$, helicity $\Lambda$, and the third components of angular momentum $J_z$, linear momentum $P_z$, and ``spin angular momentum'' $\hat{S}_z$. When the basis is chosen so that the eigenvalue of the quantity of interest is one of its defining numbers, $\Gamma$ is diagonal. For $H$, $\Lambda$ and $J_z$ we have chosen the basis of multipoles of well-defined helicity $|\omega \ j\ m\ \lambda\rangle$  (\cite[Sec. 11.4.1]{Tung1985},\cite[\S 16]{Berestetskii1982}). For $P_z$ and $\hat{S}_z$, we have used the basis of plane-waves of well-defined helicity $|\omega \ \theta\ \phi\ \lambda\rangle$, where the momentum of the plane-wave is $\mathbf{p}=\frac{\omega}{c}[\sin\theta\cos\phi,\sin\theta\sin\phi,\cos\theta]$. The incoming(outgoing) character of the multipolar basis is determined by the appearance of incoming(outgoing) spherical Hankel functions in their expressions. For plane-waves, we can isolate their incoming(outgoing) portions using the expansion of a plane-wave into regular multipolar fields. One only needs to decompose the spherical Bessel functions that appear in the regular multipoles (see \cite[Eq. 8.7-15]{Tung1985}) into their incoming (-) and outgoing (+) Hankel constituents [see e.g. Eq. (\ref{eq:sph})]:
\begin{equation}
	\label{eq:sphinout}
	j_l(\cdot)=\frac{h_l^-(\cdot)+h_l^+(\cdot)}{2}.
\end{equation}
Table \ref{tab:gammamatrices} shows the diagonal elements of $\Gamma$ for the properties that we will analyze in the example.
Finally, the last ingredient in the calculations is the representation of the incoming illumination: A single monochromatic plane-wave of helicity $+1$ (i.e. left handed polarized), vacuum wavelength equal to $222.8$ $\mu$m, and momentum aligned with the positive $z$ direction. Its expansion into the $|\omega \ \theta\ \phi\ \lambda\rangle^{\textrm{in}}$ basis is just a single 1 in the appropriate position. Its expansion into the $|\omega \ j\ m\ \lambda\rangle^{\textrm{in}}$ is identical to the expansion of a regular plane-wave into regular multipoles \cite[Eqs. 8.7-14]{Tung1985}.

\subsection{Extra implementation notes}
Most T-matrix codes use the ``electric'' and ``magnetic'' multipoles of well-defined parity $|\omega\ j\ m\ \tau\rangle$. The multipoles of well-defined helicity in Tab. \ref{tab:mat} are related to the multipoles of well-defined parity as \cite[Eq. 11.4-6]{Tung1985}:
\begin{equation}
	\label{eq:hpt}
	\sqrt{2}|\omega\ j\ m\ \tau\rangle^{\textrm{in(out)}} = |\omega\ j\ m\ +\rangle^{\textrm{in(out)}}+\tau|\omega\ j\ m\ -\rangle^{\textrm{in(out)}}.
\end{equation}
where the parity of $|\omega\ j\ m\ \tau\rangle$ is $\tau(-1)^j$ \cite[Eq. 11-4.7]{Tung1985}. The value $\tau=1$ corresponds to the ``electric'' multipoles, and $\tau=-1$ to the ``magnetic'' multipoles (\cite[Eq. 11.4-25]{Tung1985}, \cite[p. 18]{Berestetskii1982}). 

For computing the transfer of the angular momentum component in an arbitrary direction $\mathbf{\hat{t}}$, the change of basis between multipoles of well-defined $J_z$ and multipoles of well-defined $J_t$ can be obtained by the corresponding rotation matrix (see the last part of \cite[App. A]{FerCor2016b} for the case $\mathbf{\hat{t}}=\mathbf{\hat{y}}$). In that basis, the form of $\Gamma=J_t$ is the same as the form of $\Gamma=J_z$ in Tab. \ref{tab:gammamatrices}.

For computing the transfer of the linear momentum component in an arbitrary direction $\mathbf{\hat{t}}$ the change of basis between multipoles of well-defined $J_z$ and plane-waves with a well-defined $P_t$ can be achieved in two steps. First, going to a basis of well-defined $J_t$ as indicated above, and then changing from  well-defined $J_t$ to the plane-waves with well-define $P_t$. This last transformation is identical to that between multipolar fields of well-defined $J_z$ and helicity, and plane-waves of well defined helicity and $P_x$, $P_y$, $P_z$. It can be implemented using \cite[Eqs. 8.1-11, 9.4-8]{Tung1985} as written in \cite[Eqs. A6-A11]{FerCor2016b}. Again, the form of the $\Gamma=P_t$ matrix in this basis is given by that of $\Gamma=P_z$ in Tab. \ref{tab:gammamatrices}. The treatment of the component of $\mathbf{\hat{S}}$ in an arbitrary direction follows the same pattern.
\end{document}